\documentclass[aps,preprint]{revtex4}%
\usepackage{amsfonts}
\usepackage{amsmath}
\usepackage{amssymb}
\usepackage{graphicx}
\usepackage{color}
\usepackage{hyperref}%
\providecommand{\U}[1]{\protect\rule{.1in}{.1in}}

\begin{document}
\preprint{ }
\title[ ]{Entanglement of Tripartite States with Decoherence in Noninertial frames}
\author{Salman Khan}
\email{sksafi@phys.qau.edu.pk}
\affiliation{Department of Physics, Quaid-i-Azam University, Islamabad 45320, Pakistan}
\date{Jun 8 2011}

\begin{abstract}
The one-tangle and $\pi$-tangle are used to quantify the entanglement of a
tripartite \textit{GHZ} state in noninertial frames when the system interacts
with a noisy environment in the form of phase damping, phase flip and bit flip
channel. It is shown that the two-tangles behave as a closed system. The
one-tangle and $\pi$-tangle have different behaviors in the three channel. In
the case of phase damping channel, depending on the kind of coupling, the
sudden death of both one-tangle and $\pi$-tangle may or may not happen.
Whereas in the case of phase flip channel the sudden death cannot be avoided.
The effect of decoherence may be ignored in the limit of infinite acceleration
when the system interacts with a bit flip channel. Furthermore, a sudden
rebirth of the one-tangle and $\pi$-tangle occur in the case of phase flip
channel that may be delayed when collective coupling is switched on.



\end{abstract}
\maketitle

\section{Introduction}

Entanglement is not only one of the most striking properties of quantum
mechanics but also the core concept of quantum information and quantum
computation \cite{springer}. The structure of all the subfields of quantum
information theory, such as teleportation of unknown states \cite{Bennett},
quantum key distribution \cite{Ekert}, quantum cryptography \cite{Bennett2}
and quantum computation \cite{Grover, Vincenzo}, stand on quantum
entanglement. The dynamics of entanglement in various bipartite qubit and
qutrit states have been extensively studied under various circumstances. The
study of entanglement of various fields in the accelerated frames has taken
into account recently and valuable results about the behavior of entanglement
have been obtained \cite{Alsing,Ling,Gingrich,Pan, Schuller, Terashima}. The
effect of decoherence on the behavior of entanglement under various quantum
channels in noninertial frames have been studied in Refs. \cite{Wang, Salman}.
However, these studies are limited only to bipartite qubit systems in the
accelerated frames. Recently, the dynamics of entanglement in tripartite qubit
systems in noninertial frames are studied in Refs. \cite{Hwang, Wang2,
Shamirzai}. These studies show that the degree of entanglement is degraded by
the acceleration of the frames and, like the two-tangles in inertial frames,
the two-tangles are zero when one or two observers are in the accelerated frames.

In this paper, I investigate the effect of decoherence on the tripartite
entanglement of Dirac field in accelerated frames by using a phase damping
channel, a phase flip channel and a bit flip channel. The effect of amplitude
damping channel and depolarizing channel on tripartite entanglement of Dirac
field in a noninertial frames is recently studied in Ref. \cite{Winpin}. I
consider three observers Alice, Bob and Charlie that initially share a
\textit{GHZ} tripartite state. One of the observers, say Alice, stays
stationary and the other two observers move with a constant acceleration. I
work out the effect of acceleration and of decoherence by using the three
channels on the initial entanglement of the shared state between the
observers. I consider different kinds of coupling of each channel with the
system. For example, in one case each qubit interacts locally with the noisy
environment. In the second case, all the three qubits are influenced
collectively by the same environment. I show that the entanglement sudden
death (ESD) \cite{Yu1, Yu2} can either be completely avoided or can be slowed
down depending on the various coupling of the system and a particular channel.
Furthermore, I also show that the ESD can happen faster and becomes
independent of the acceleration when the system interacts with a phase flip channel.%

\begin{table*}[htb]%
\caption{A single qubit Kraus operators for phase damping channel, phase
flip channel and bit flip channel. \label{table:1}}%
\begin{tabular}
[c]{|c|c|}\hline
phase damping channel & $E_{o}=\left(
\begin{array}
[c]{cc}%
1 & 0\\
0 & \sqrt{1-p}%
\end{array}
\right)  ,\qquad E_{1}=\left(
\begin{array}
[c]{cc}%
0 & 0\\
0 & \sqrt{p}%
\end{array}
\right)  $\\\hline
phase flip channel & $E_{o}=\sqrt{1-p}\left(
\begin{array}
[c]{cc}%
1 & 0\\
0 & 1
\end{array}
\right)  ,\qquad E_{1}=\sqrt{p}\left(
\begin{array}
[c]{cc}%
1 & 0\\
0 & -1
\end{array}
\right)  $\\\hline
bit flip channel & $E_{o}=\sqrt{1-p}\left(
\begin{array}
[c]{cc}%
1 & 0\\
0 & 1
\end{array}
\right)  ,\qquad E_{1}=\sqrt{p}\left(
\begin{array}
[c]{cc}%
0 & 1\\
1 & 0
\end{array}
\right)  $\\\hline
\end{tabular}
%

\end{table*}%

\section{The system in noisy environment}

The evolution of a density matrix of a system in a noisy environment is
described in terms of Kraus operators formalism. The Kraus operators for a
single qubit system of the three channels that I use in this paper are given
in Table $1$. Each channel is parameterized by the decoherence parameter $p$
that has values between $0$ and $1$. For the lower limit of $p$, a channel has
no effect on the system and is said to be undecohered whereas for the upper
limit it introduces maximum error and is said to be fully decohered. The final
density matrix of a composite system when it evolves in a noisy environment is
given by the following equation%
\begin{equation}
\rho_{f}=\sum_{i,j,k,...}K_{i}K_{j}K_{k}...\rho...K_{k}^{\dag}K_{j}^{\dag
}K_{i}^{\dag}, \label{1}%
\end{equation}
where $\rho$ is the initial density matrix of the system and $K_{n}$ are the
Kraus operators that satisfy the completeness relation $\sum_{n}K_{n}^{\dag
}K_{n}=I$. I consider that Alice, Bob and Charlie initially share the
following maximally entangled \textit{GHZ} state%
\begin{equation}
|\psi\rangle_{ABC}=\frac{1}{\sqrt{2}}\left(  |000\rangle_{A,B,C}%
+|111\rangle_{A,B,C}\right)  , \label{2}%
\end{equation}
where the three alphabets represent the three observers and the entries in
each ket correspond to the three observers in the order of alphabet. From the
perspective of an inertial observer, the Dirac fields can be expressed as a
superposition of Minkowski monochromatic modes $|0\rangle_{M}=\bigoplus
_{i}|0_{\omega_{i}}\rangle_{M}$ and $|1\rangle_{M}=\bigoplus_{i}|1_{\omega
_{i}}\rangle_{M}$ $\forall i$, with \cite{AsPach, Martin}
\begin{align}
|0_{\omega_{i}}\rangle_{M}  &  =\cos r_{i}|0_{\omega_{i}}\rangle_{I}%
|0_{\omega_{i}}\rangle_{II}+\sin r_{i}|0_{\omega_{i}}\rangle_{I}|0_{\omega
_{i}}\rangle_{II},\nonumber\\
|1_{\omega_{i}}\rangle_{M}  &  =|1_{\omega_{i}}\rangle_{I}|1_{\omega_{i}%
}\rangle_{II}, \label{3}%
\end{align}
where $\cos r_{i}=\left(  e^{-2\pi\omega_{i}c/a_{i}}+1\right)  ^{-1/2}$. The
parameters $\omega_{i}$, $c$ and $a_{i}$, in the exponential stand,
respectively, for Dirac particle's frequency, speed of light in vacuum and
acceleration of the $i$th observer. In Eq. (\ref{3}) the subscripts $I$ and
$II$ of the kets represent the Rindler modes in region $I$ and $II$,
respectively, in the Rindler spacetime diagram. The Minkowski mode that
defines the Minkowski vacuum is related to a highly nonmonochromatic Rindler
mode rather than a single mode with the same frequency \cite{Martin, Bruschi}.
Let the angular frequencies $\omega_{i}$ for the three observers be,
respectively, given by $\omega_{a}$, $\omega_{b}$ and $\omega_{c}$ then with
respect to an inertial observer $|0_{\omega_{a(b,c)}}\rangle_{A(B,C)}$ and
$|1_{\omega_{a(b,c)}}\rangle_{A(B,C)}$ are the vacuum and the first excited
states. Furthermore, let us suppose that each observer is equipped with a
monochromatic device that are sensitive only to their respective modes. In
order to save space and present the relation in simple form, I drop the
frequencies in subsript of each entry of the kets. Then substituting the
Rindler modes from Eq. (\ref{3}) for the Minkowski modes in Eq. (\ref{2}) for
the two observers in the accelerated frames gives%
\begin{align}
|\psi\rangle_{ABC}  &  =\frac{1}{\sqrt{2}}(\cos r_{b}\cos r_{c}|00000\rangle
_{A,BI,BII,CI,CII}\nonumber\\
&  +\cos r_{b}\sin r_{c}|00011\rangle_{A,BI,BII,CI,CII}\nonumber\\
&  +\sin r_{b}\cos r_{c}|01100\rangle_{A,BI,BII,CI,CII}\nonumber\\
&  +\sin r_{b}\sin r_{c}|01111\rangle_{A,BI,BII,CI,CII}\nonumber\\
&  +|11010\rangle_{A,BI,BII,CI,CII}). \label{4}%
\end{align}
Let the noninertial frames of Bob and Charlie be moving with same
acceleration, then $r_{b}=r_{c}=r$. Since the Rindler modes in region $II$ are
inaccessible, tracing out over those modes leave the following initial mixed
density matrix%
\begin{align}
\rho &  =\frac{1}{2}[\cos^{2}r_{b}\cos^{2}r_{c}|000\rangle\langle000|+\cos
^{2}r_{b}\sin^{2}r_{c}|001\rangle\langle001|\nonumber\\
&  +\sin^{2}r_{b}\cos^{2}r_{c}|010\rangle\langle010|+\sin^{2}r_{b}\sin
^{2}r_{c}|011\rangle\langle011|\nonumber\\
&  +\cos r_{b}\cos r_{c}(|000\rangle\langle111|+|111\rangle\langle
000|)+|111\rangle\langle111|. \label{5}%
\end{align}
Note that I have dropped the subsript $I$ that indicates the Rindler modes in
region $I$. In the rest of the paper, all calculations correspond to the
Rindler modes in region $I$.

The degree of entanglement in bipartite qubit states is quantified by using
the concept of negativity \cite{Peres, Horodecki1}. It is given by%
\begin{equation}
\mathcal{N}_{AB}=\left\Vert \rho_{AB}^{T_{B}}\right\Vert -1, \label{6}%
\end{equation}
where $T_{B}$ is the partial transpose over the second qubit $B$ and
$\left\Vert .\right\Vert $ gives the trace norm of a matrix. For a bipartite
density matrix $\rho_{m\nu,n\mu}$, the partial transpose over the second qubit
$B$ is given by $\rho_{m\mu,n\nu}^{T_{B}}=\rho_{m\nu,n\mu}$ and for the first
qubit, it can similarly be defined. On the other hand, the entanglement of a
tripartite state $|\psi\rangle_{ABC}$ is quantified by using the concept of
$\pi$-tangle, which is given by%
\begin{equation}
\pi_{ABC}=\frac{1}{3}(\pi_{A}+\pi_{B}+\pi_{C}), \label{7}%
\end{equation}
where $\pi_{A(BC)}$ is the residual entanglement and is given by%
\begin{equation}
\pi_{A}=\mathcal{N}_{A(BC)}^{2}-\mathcal{N}_{AB}^{2}-\mathcal{N}_{AC}^{2}.
\label{8}%
\end{equation}
In Eq. (\ref{8}), $\mathcal{N}_{AB(AC)}$ is a two-tangle and is given as the
negativity of the mixed density matrix $\rho_{AB(AC)}=Tr_{C(B)}|\psi
\rangle_{ABC}\langle\psi|$ whereas $\mathcal{N}_{A(BC)}$ is a one-tangle and
is defined as $\mathcal{N}_{A(BC)}=\left\Vert \rho_{ABC}^{T_{A}}\right\Vert
-1$.

Now, in the following I use the three channels to investigate the behavior of
entanglement of the tripartite initially entangled density matrix of Eq.
(\ref{8}) in a noisy environment.

\subsection{Phase damping channel}

When all the three qubits are locally coupled to a phase damping channel, then
Eq. (\ref{1}) for the evolution of a tripartite state can be written as%
\begin{equation}
\rho_{f}=\sum_{i,j,k}(K_{i}K_{j}K_{k})\rho\left(  K_{k}^{\dag}K_{j}^{\dag
}K_{i}^{\dag}\right)  . \label{9}%
\end{equation}
The Kraus operators in Eq. (\ref{9}) with subscripts $i$, $j$, $k$ act,
respectively, on Alice's, Bob's and Charlie's qubits. These operators are
formed from the single qubit Kraus operators of a phase damping channel and
are, respectively, given by $K_{i}=(E_{i}\otimes I\otimes I)$, $K_{j}%
=(I\otimes E_{j}\otimes I)$, $K_{k}=(I\otimes I\otimes E_{k})$ with $i=j=k=0$,
$1$. Using the initial density matrix of Eq. (\ref{5}) in Eq. (\ref{9}), the
final density matrix of the system is given as%
\begin{align}
\rho_{f}  &  =\frac{1}{2}\cos^{4}r|000\rangle\langle000|+\frac{1}{8}\sin
^{2}2r(|001\rangle\langle001|+|010\rangle\langle010|)\nonumber\\
&  +\frac{1}{2}\sin^{4}r|011\rangle\langle011|+\frac{1}{2}|111\rangle
\langle111|\nonumber\\
&  +\frac{1}{2}\sqrt{(1-p_{0})(1-p_{1})(1-p_{2})}\cos^{2}r(|000\rangle
\langle111|+|111\rangle\langle000|), \label{10}%
\end{align}
where $p_{0}$, $p_{1}$ and $p_{2}$ are the decoherence parameters of the
locally coupled channels with the qubits of Alice, Bob and Charlie,
respectively. The three one-tangles that can straightforwardly be calculated
by using its definition are given as%
\begin{align}
\mathcal{N}_{A(BC)}  &  =\frac{1}{4}(-2+2\cos^{4}r+2\sqrt{(1-p_{0}%
)(1-p_{1})(1-p_{2})\cos^{4}r}\nonumber\\
&  +2\sqrt{(1-p_{0})(1-p_{1})(1-p_{2})\cos^{4}r+\sin^{8}r}+\sin^{2}2r),
\label{11}%
\end{align}%
\begin{align}
\mathcal{N}_{B(AC)}  &  =\mathcal{N}_{C(AB)}=\frac{1}{16}(-1+8\sqrt
{(1-p_{0})(1-p_{1})(1-p_{2})\cos^{4}r}\nonumber\\
&  +\cos4r+2\sqrt{16(1-p_{0})(1-p_{1})(1-p_{2})\cos^{4}r+\sin^{4}2r}).
\label{12}%
\end{align}

\begin{figure}[h]
\begin{center}%
\begin{tabular}
[c]{cc}%
\includegraphics[scale=1.2]{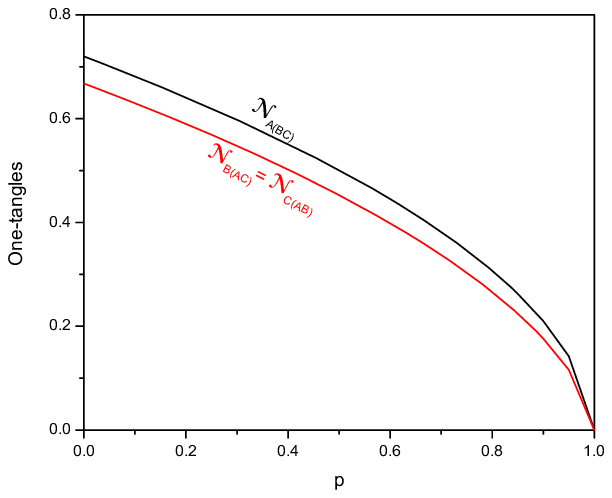} \put( -480,230) \ \ (a) &
\end{tabular}
\ \ \ \ \ \ \ \ \ \ \ \ \ \ \ \ \ \ \ \ \ \ \ \ \ \
\begin{tabular}
[c]{cc}%
\includegraphics[scale=1.2]{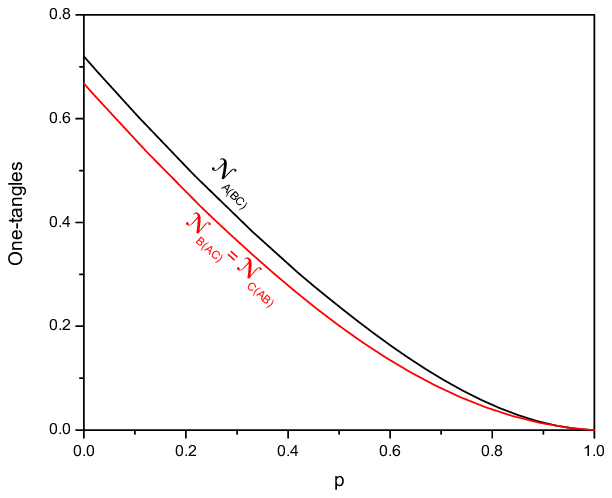} \put(-480,230) \ \ \ (b) &
\end{tabular}
\end{center}
\caption{(color online) (a) The one-tangles, for $r=\pi/6$, are plotted
against the decoherence parameter $p$ when only one qubit is coupled to the
phase damping channel. (b) The one tangles, for $r=\pi/6$, are plotted against
$p$ when all the three qubit are influenced collectively by the same phase
damping channel.}%
\label{Figure1ab}%
\end{figure}Note that $\mathcal{N}_{B(AC)}=\mathcal{N}_{C(AB)}$ shows that the
two accelerated subsystems are symmetrical for any value of the acceleration.
It can easily be checked that all the one-tangles reduce to $[(1-p_{0}%
)(1-p_{1})(1-p_{2})]^{1/2}$ when $r=0$ and for $p_{i}=r=0,$ the result is $1$,
which is the result obtained in the rest frames both for Dirac and scalar
fields \cite{Hwang, Wang2}. It is also clear that for $r=0$ case, the
one-tangles goes to zero when either of the three qubits is coupled to a fully
decohered channel. Moreover, for $r=\pi/4$ all the three one-tangles become
indistinguishable, that is, $\mathcal{N}_{A(BC)}=\mathcal{N}_{B(AC)}%
=\mathcal{N}_{C(AB)}$ irrespective of the value of decoherence parameter and
of which qubit is coupled to the channel. Also, in this case the one-tangle is
lost only when either of the three local channels is fully decohered. However,
the initial entanglement is small enough as compared to the case for $r=0$.
For example, when only one qubit is locally coupled to the channel, the
initial value of the one-tangle is $0.4$ for $p_{i}=0$. This shows the strong
dependence of the one tangle over the acceleration. However, the on-tangles
are different for other values of the acceleration. To see the behavior of the
one-tangles for other values of the acceleration, I plot it against the
decoherence parameters for $r=\pi/6$ in Fig. $1$($a,b$). Fig. $1a$ shows the
behavior of the one-tangles when only one qubit is locally influenced by the
channel. This behavior does not change by switching the coupling of the
channel from one qubit to another, that is, whether it's plotted against
$p_{0}$ or $p_{1}$($p_{2}$), which means that the one-tangles are symmetrical
in the phase damping environment. It can be seen from the figure that both
$\mathcal{N}_{A(BC)}$ and $\mathcal{N}_{B(AC)}$($\mathcal{N}_{C(AB)}$) varies
identically with the increasing value of decoherence parameter, however, the
acceleration effect is different on both of them. Like for the case of $r=0$
and $r=\pi/4$, the one-tangles goes to zero only when the channel is fully
decohered, which shows that no sudden death occurs. The behavior of
one-tangles when all the three qubits are under the influence of collective
environment ($p_{0}=p_{1}=p_{2}=p$) is shown in Fig. $1b$ for $r=\pi/6$. In
this case, the one-tangles are heavily damped by the decoherence parameter as
compared to Fig $1a$ and become indistinguishable at $p=0.85$.

Next, I find the two-tangles according to the definition given above. Taking
partial trace of the final density matrix of Eq. (\ref{10}) over the Bob's
qubit or Charlie's qubit leads to the following mixed state%
\begin{equation}
\rho_{AB(AC)}=\rho_{ABC}^{T_{C(B)}}=\frac{1}{2}(\cos^{2}r|00\rangle
\langle00|+\sin^{2}r|01\rangle\langle01|+|11\rangle\langle11|). \label{13}%
\end{equation}
Since non of the elements of the density matrix depends on the decoherence
parameter, the two-tangles are unaffected by the noisy environment. Similarly,
one can straightforwardly prove that tracing over Alice's qubit leads to a
density matrix whose every element is independent of the decoherence
parameter. From these matrices it can easily be verified that $\mathcal{N}%
_{AB}=\mathcal{N}_{AC}=\mathcal{N}_{BC}=0$. The zero value of all the
two-tangles verifies that no entanglement exists between any two subsystems of
the tripartite state.

\begin{figure}[h]
\begin{center}%
\begin{tabular}
[c]{ccc}%
\vspace{-0.5cm} \includegraphics[scale=1.2]{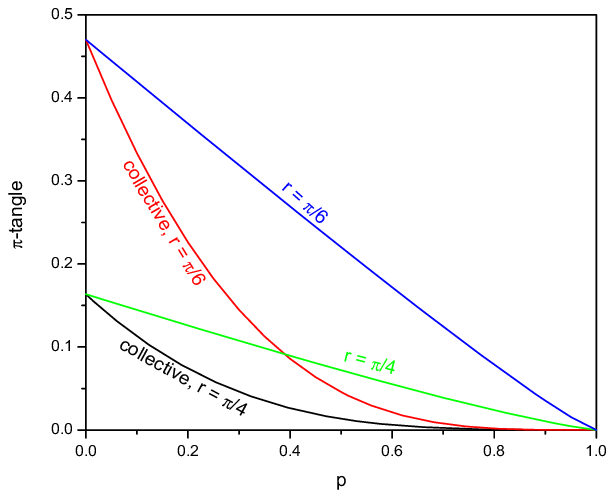}\put(-320,220) &  &
\end{tabular}
\end{center}
\caption{(color online) The $\pi$-tangle is plotted against the decoherence
parameter for $r=\pi/6$ and $r=\pi/4$ both for one qubit coupled with the
phase damping channel and for all the three qubit coupled collectively with
the same phase damping channel.}%
\label{Figure2}%
\end{figure}

I am now in position to find the $\pi$-tangle by using its defining Eq.
(\ref{7}). Since all the two-tangles are zero, the $\pi$-tangle simply becomes%
\begin{align}
\pi_{ABC}  &  =\frac{1}{3}\left(  \mathcal{N}_{A(BC)}^{2}+\mathcal{N}%
_{B(AC)}^{2}+\mathcal{N}_{C(AB)}^{2}\right) \nonumber\\
&  =\frac{1}{384}[8(-2+2\cos^{4}r+2\sqrt{(1-p_{0})(1-p_{1})(1-p_{2})\cos^{4}%
r}\nonumber\\
&  +2\sqrt{(1-p_{0})(1-p_{1})(1-p_{2})\cos^{4}r+\sin^{8}r}+\sin^{2}%
2r)^{2}\nonumber\\
&  +(-1+8\sqrt{(1-p_{0})(1-p_{1})(1-p_{2})\cos^{4}r}+\cos4r\nonumber\\
&  +2\sqrt{16(1-p_{0})(1-p_{1})(1-p_{2})\cos^{4}r+\sin^{4}2r})^{2}].
\label{14}%
\end{align}
From Eq. (\ref{14}), it can easily be verified that for the case of inertial
frames the $\pi$-tangle reduces to $(1-p_{i})$ when only one qubit is locally
coupled to the noisy environment. This is easy to check that this result goes
to zero only when the decoherence parameter is $1$. Moreover, for any
acceleration, the $\pi$-tangle does not depend on whether the stationary or
the accelerated qubit is coupled to the environment. In Fig. $2$, I plot the
$\pi$-tangle for different accelerations when only one qubit is coupled to the
environment. It can be seen that the $\pi$-tangle is lost only when the
channel is fully decohered and thus no sudden death occurs in this case as
well. I have also plotted the $\pi$-tangle for the case of collective
environment in the same figure. The graphs show that the sudden death can
happen when the system is influenced by collective environment.

\subsection{Phase flip channel}

When the three qubits are locally coupled to a phase flip channel, the final
density matrix of the system is again given by Eq. (\ref{9}). The subscripts
and the number of Kraus operators are the same as described for the previous
case, however, these are now made from single qubit phase flip Kraus
operators. The final density matrix of the system in this case becomes%
\begin{align}
\rho_{f}  &  =\frac{1}{2}\cos^{4}r|000\rangle\langle000|+\frac{1}{8}\sin
^{2}2r(|001\rangle\langle001|+|010\rangle\langle010|)\nonumber\\
&  +\frac{1}{2}(1-2p_{0})(1-2p_{1})(1-2p_{2})\cos^{2}r(|000\rangle
\langle111|+|111\rangle\langle000|)\nonumber\\
&  +\frac{1}{2}\sin^{4}r|011\rangle\langle011|+\frac{1}{2}|111\rangle
\langle111|. \label{15a}%
\end{align}
Using the definition of one-tangle, the three one-tangles in this case become%
\begin{align}
\mathcal{N}_{A(BC)}  &  =1/4(-2+2\cos^{2}r(\left\vert (1-2p_{0})(1-2p_{1}%
)(1-2p_{2})\right\vert +\cos^{2}r)\nonumber\\
&  +2\sqrt{(1-2p_{0})^{2}(1-2p_{1})^{2}(1-2p_{2})^{2}\cos^{4}r+\sin^{8}r}%
+\sin^{2}2r)\nonumber\\
\mathcal{N}_{B(AC)}  &  =\mathcal{N}_{C(AB)}=\frac{1}{8}(-4+4\left\vert
(1-2p_{0})(1-2p_{1})(1-2p_{2})\right\vert \cos^{2}r+4\cos^{4}r\nonumber\\
&  +4\sin^{4}r+\sin^{2}2r+\sqrt{16(1-2p_{0})^{2}(1-2p_{1})^{2}(1-2p_{2}%
)^{2}\cos^{4}r+\sin^{4}2r}) \label{16}%
\end{align}

\begin{figure}[h]
\begin{center}%
\begin{tabular}
[c]{cc}%
\includegraphics[scale=1.2]{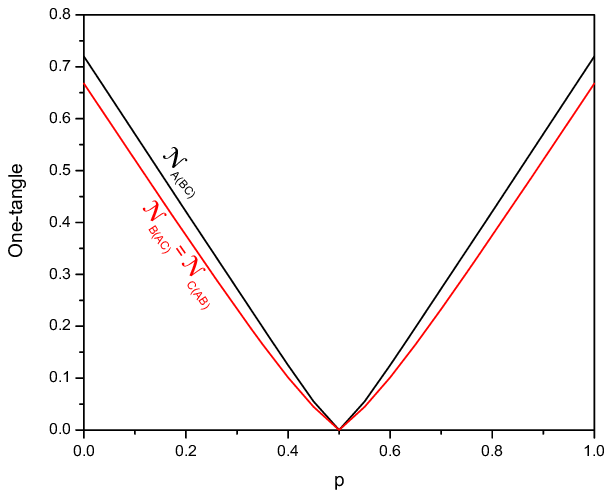} \put( -480,230) \ \ (a) &
\end{tabular}
\ \ \ \ \ \ \ \ \ \ \ \ \ \ \ \ \ \ \ \ \ \ \ \ \ \ \ \ \
\begin{tabular}
[c]{cc}%
\includegraphics[scale=1.2]{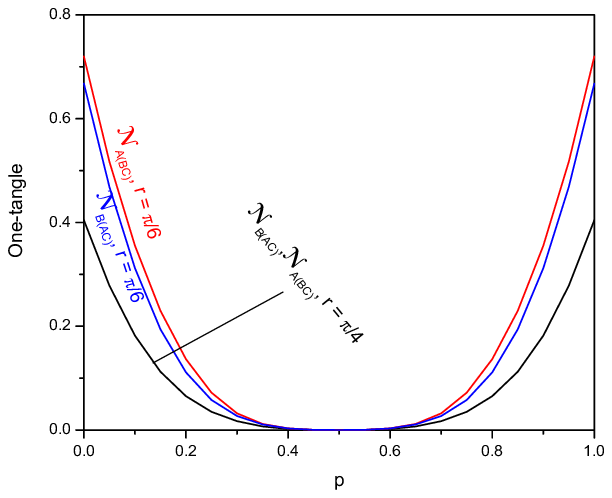} \put(-480,230) \ \ \ (b) &
\end{tabular}
\end{center}
\caption{(color online) (a) The one-tangles, for $r=\pi/6$, are plotted
against the decoherence parameter $p$ when only one qubit is coupled to the
phase flip channel. (b) The one tangles, for $r=\pi/6$, are plotted against
$p$ when all the three qubit are influenced collectively by the\ phase flip
channel.}%
\label{Figure3}%
\end{figure}where, $\left\vert .\right\vert $ represents the absolute value.
Once again, the equality between the one-tangles for the accelerated frames
shows the symmetrical nature of the problem. Also, for $r=0$, the three
one-tangles are indistinguishable, as expected, and are given by $\left\vert
(1-2p_{0})(1-2p_{1})(1-2p_{2})\right\vert $. This shows that the one-tangles
vanishes at $p_{i}=0.5$ irrespective of which qubit is locally coupled to the
noisy environment. For undecohered case, this leads to the result of static
frames. Again, like in the case of phase damping channel, all the one-tangles
becomes indistinguishable at $r=\pi/4$, however, unlike the phase damping
channel it goes to zero at $p_{i}=0.5$. This means that the phase flip channel
destroys all the one-tangles earlier, in other words, the tripartite
entanglement is caused to sudden death. The one-tangles are shown in Fig. $3a$
for $r=\pi/6$ when only one qubit is locally influenced by the phase flip
channel. It is important to mention here that the channel effect is locally
symmetrical for both inertial and noninertial qubits. One can see that
initially the one-tangles are acceleration dependent and this dependence on
acceleration gradually decreases as the decoherence parameter increases and
the difference vanishes near $p_{i}=0.5$. This shows that the one-tangles are
heavily damped by the decoherence parameter in comparison to the damping
caused by the acceleration. Hence one can ignore the acceleration effect near
a $50\%$ decoherence level. All the one-tangles become zero at $p_{i}=0.5$,
which in fact happens for any acceleration. Interestingly enough, the
one-tangles' sudden rebirth takes place for values of $p_{i}>0.5$ irrespective
of the acceleration of the frames. However, the regrowing rate is different
for different values of acceleration. This means that at $50\%$ decoherence
level all the tripartite entanglement first shifts to the environment and then
part of it shifts back to the system at values of decoherence parameter higher
than this value. In Fig. $3b$, the effect of collective environment on all the
one-tangles is shown for $r=\pi/6$ and $r=\pi/4$. It can be seen that the
damping caused by the collective environment is heavier than the damping in
one qubit local coupling case and a sudden death prior to a $50\%$ decoherence
level may happen. Moreover, the sudden rebirth of the one-tangles is also delayed.

One can further verify that by taking partial trace of Eq. (\ref{15a}) over
Bob's or Charlie's qubit leads to the same result as given in Eq. (\ref{13}).
This means that no bipartite entanglement exits for this case as well, because
all the two-tangles are zero as before. The $\pi$-tangle can be found
straightforwardly as done in the previous case, which is one-third of the sum
of the squares of the three one-tangles. It is given by%
\begin{align}
\pi_{ABC}  &  =\frac{1}{96}[2(-2+2\cos^{2}r(\left\vert (1-2p_{0}%
)(1-2p_{1})(1-2p_{2})\right\vert +\cos^{2}r)\nonumber\\
&  +2\sqrt{(1-2p_{0})^{2}(1-2p_{1})^{2}(1-2p_{2})^{2}\cos^{4}r+\sin^{8}r}%
+\sin^{2}2r)^{2}\nonumber\\
&  +(-4+4\left\vert 1-2p_{0}\right\vert \left\vert 1-2p_{1}\right\vert
\left\vert 1-2p_{2}\right\vert \cos^{2}r+4\cos^{4}r+4\sin^{4}r\nonumber\\
&  +\sin^{2}2r+\sqrt{16(1-2p_{0})^{2}(1-2p_{1})^{2}(1-2p_{2})^{2}\cos
^{4}r+\sin^{4}2r})^{2}].
\end{align}

\begin{figure}[h]
\begin{center}%
\begin{tabular}
[c]{ccc}%
\vspace{-0.5cm} \includegraphics[scale=1.2]{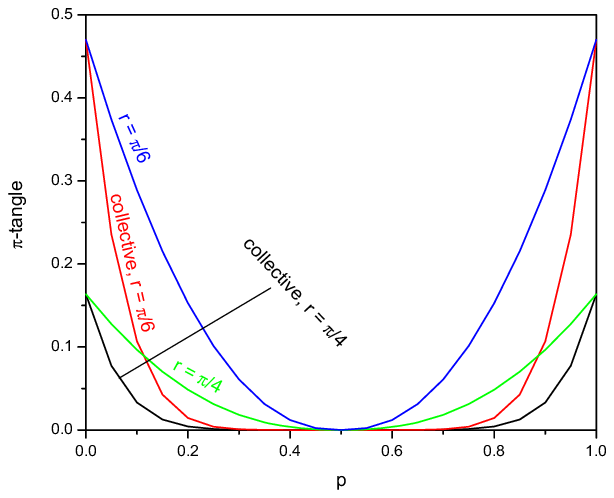}\put(-320,220) &  &
\end{tabular}
\end{center}
\caption{(color online) The $\pi$-tangle is plotted against the decoherence
parameter for $r=\pi/6$ and $r=\pi/4$ both for one qubit coupled with the
phase flip channel and for all the three qubit coupled collectively with the
same phase flip channel.}%
\label{Figure4}%
\end{figure}I plot it for one qubit local coupling case for $r=\pi/6$, $\pi/4$
in Fig. $4$. It can be seen that the $\pi$-tangle is strongly acceleration
dependent for the lower limit of the decoherence parameter. However, as the
decoherence parameter increases, the acceleration dependence decreases and it
goes to zero, irrespective of the value of acceleration, at $p_{i}=0.5$. The
sudden rebirth of the $\pi$-tangle is not as quick as in the case of
one-tangles. This behavior of the $\pi$-tangle does not depend on which one
qubit is coupled to the environment. I have also plotted the $\pi$-tangle for
the case of collective environment for $r=\pi/6$, $\pi/4$. The figure shows
that the sudden death is faster when the system evolves under collective
environment and the rebirth is quite delayed.

\begin{figure}[h]
\begin{center}%
\begin{tabular}
[c]{cc}%
\includegraphics[scale=1.2]{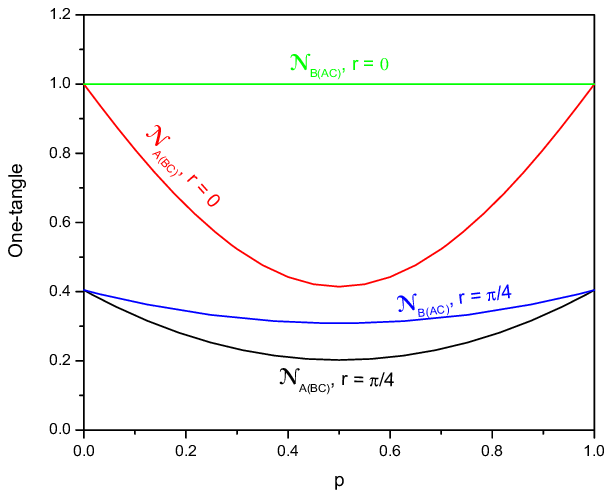} \put( -480,230) \ \ (a) &
\end{tabular}
\ \ \ \ \ \ \ \ \ \ \ \ \ \ \ \ \ \ \ \ \ \ \ \ \ \ \ \ \
\begin{tabular}
[c]{cc}%
\includegraphics[scale=1.2]{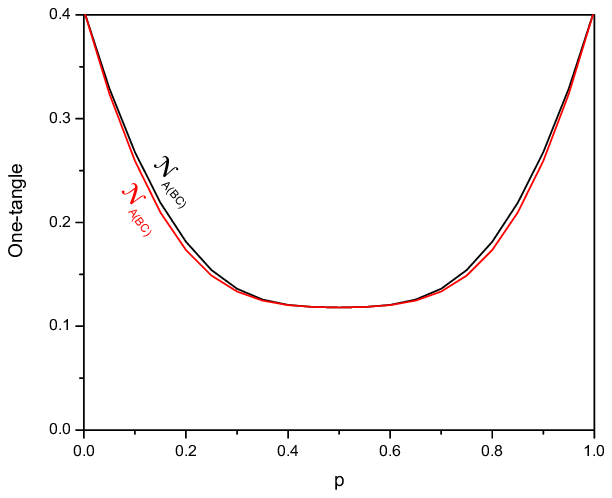} \put(-480,230) \ \ \ (b) &
\end{tabular}
\end{center}
\caption{(color online) (a) The one-tangles, for $r=0$ and $r=\pi/4$, are
plotted against the decoherence parameter $p$ when only Alice's qubit is
coupled to the bit flip channel. (b) For the same two values of acceleration,
the one-tangles are plotted against the decoherence parameter when the
collective environment is switched on.}%
\label{Figure5}%
\end{figure}

\subsection{Bit flip channel}

Finally, I investigate the effect of bit flip channel on the entanglement of
the tripartite system. When each qubit is locally coupled to the channel, the
final density matrix of the system is again given by Eq. (\ref{9}) and the
Kraus operators have the same meaning as before. However, these are now made
from the single qubit bit flip channel in the same way as mentioned earlier.
Instead of writing the mathematical relations for one-tangles, which are quite
lengthy, I prefer to see the effect of decoherence only for special cases.
Unlike the two previous cases, the one-tangles for the accelerated observers
are not always equal. Similarly, for $r=0$ case, the equality of one-tangles
depends on which qubit is locally coupled to the noisy environment. For
example, when only Alice's qubit is locally coupled to the channel, then
$\mathcal{N}_{B(AC)}=\mathcal{N}_{C(AB)}=1$ and $\mathcal{N}_{A(BC)}%
=-1+2\sqrt{1-2p_{0}+2p_{0}^{2}}$ and if Bob's qubit is under the influence of
the channel then $\mathcal{N}_{A(BC)}=\mathcal{N}_{C(AB)}=1$ and
$\mathcal{N}_{B(AC)}=-1+2\sqrt{1-2p_{1}+2p_{1}^{2}}$. A similar relation
exists when only Charlie's qubit is coupled to the environment. On the other
hand, all the three one-tangles are equal when the system is coupled to
collective environment and are given by $-1+2\sqrt{2}\sqrt{(-1+p)^{2}p^{2}%
}+2\sqrt{1-6p+16p^{2}-20p^{3}+10p^{4}}$. A similar situation arises for
$r=\pi/4$ as well. That is, two one-tangles are always equal and like for
$r=0$ case, the equality between two one-tangles depends on which qubit is
under the action of the channel. However, unlike $r=0$ case, non of the
one-tangles is always equal to $1$, that is, every tangle is decoherence
parameter dependent. Fig. $5a$ shows the dynamics of the one-tangles for
$r=0$, $\pi/4$ when only Alice's qubit is locally coupled to the channel. The
acceleration dependence of the one-tangles can easily be seen as compared to
$r=0$ case. Also, $\mathcal{N}_{B(AC)}=\mathcal{N}_{C(AB)}$ for the whole
range of decoherence parameter. Even though the behavior of each one-tangle is
symmetrical around $50\%$ decoherence level, the damping is more for
$\mathcal{N}_{A(BC)}$. The interesting features are the symmetrical increase
beyond $50\%$ decoherence level of each one-tangle back to its initial value
for a fully decohered channel and the occurrence of no sudden death even in
the limit of infinite acceleration. The behavior of one-tangles for the case
of collective environment is shown in Fig.$5b$. Even in this case of strongly
decohered system, there is no sudden death of any one-tangle. The second
remarkable feature is the three regions where the subsystems become
indistinguishable, that is, for lower limit, intermediate range and upper
limit of the decoherence parameter. In Fig. $6$, the results of one-tangles
are plotted for $r=\pi/6$ against the decoherence parameter when Alice's qubit
interacts with the channel. Again, each one-tangle is symmetrical around
$p_{1}=0.5$, however, they are affected differently. It can be seen that there
are different particular couple of values of the decoherence parameter at
which all the three one-tangles are same and the subsystems become
indistinguishable. Interestingly, in between these points, the one-tangles of
the accelerated observers are less effected than the one-tangle of the static observer.

\begin{figure}[h]
\begin{center}%
\begin{tabular}
[c]{ccc}%
\vspace{-0.5cm} \includegraphics[scale=1.2]{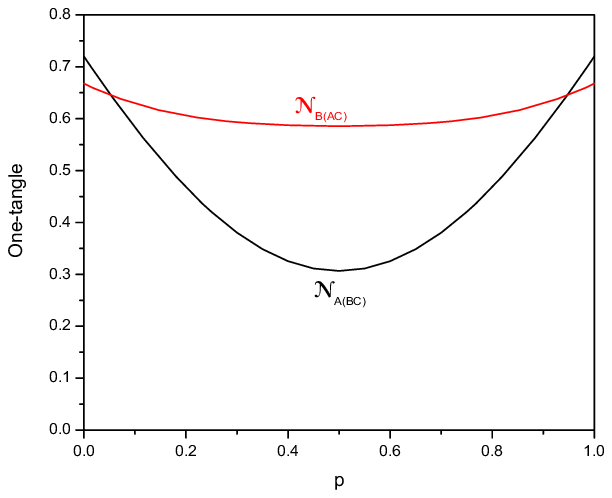}\put(-320,220) &  &
\end{tabular}
\end{center}
\caption{(color online) The one-tangles are plotted against the decoherence
parameter for $r=\pi/6$ when only Alice's qubit is locally coupled to the bit
flip channel.}%
\label{Figure6}%
\end{figure}

Like in the previous two cases, all the two-tangles are zero in this case as
well. The $\pi$-tangle can be found similar to the previous two cases. To see
how it is effected by the noisy environment, I plot it in Fig. $7$ against the
decoherence parameter for different values of the acceleration. The figure
shows the behavior of $\pi$-tangle for the case when only one qubit is
influenced by the channel. This behavior of the $\pi$-tangle does not change
by switching the coupling of the channel from one qubit to another, which
shows that the $\pi$-tangle is invariant with respect to the local coupling of
the channel with a single qubit. The acceleration dependence of the $\pi
$-tangle is quite clear from the figure. It can also be seen that the damping
caused by the decoherence parameter is heavy for small values of the
acceleration as compared to the case of large values of the acceleration. In
other words, the effect of decoherence diminishes as the acceleration
increases. This means that in the limit of infinite acceleration the
decoherence effect may be ignored. The behavior of $\pi$-tangle when the
system is under the action of collective environment has also been plotted in
Fig. $7$ for two different values of the acceleration. In comparison to single
qubit coupling, the damping is heavier, however, no sudden death occurs.

\begin{figure}[h]
\begin{center}%
\begin{tabular}
[c]{ccc}%
\vspace{-0.5cm} \includegraphics[scale=1.2]{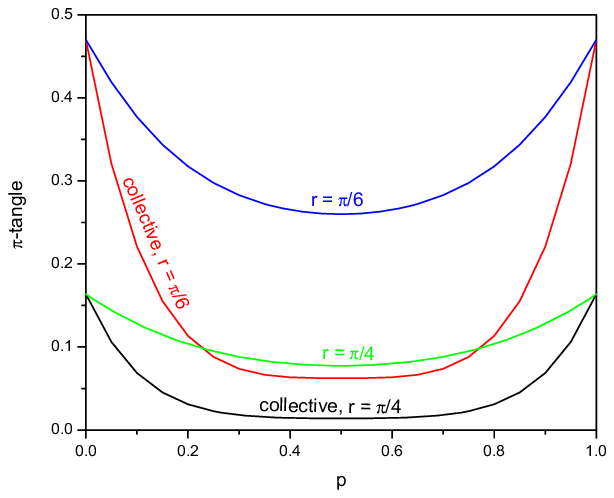}\put(-320,220) &  &
\end{tabular}
\end{center}
\caption{(color online) The $\pi$-tangle is plotted against the decoherence
parameter for $r=\pi/6$ and $r=\pi/4$ both for one (Alice) qubit coupled with
the bit flip channel and for all the three qubit coupled collectively with the
same bit flip channel.}%
\label{Figure7}%
\end{figure}

\section{Summary}

In conclusion, the effects of different channels on the tripartite
entanglement of \textit{GHZ} state in noninertial frames by using one-tangle
and $\pi$-tangle as the entanglement quantifier are investigated. It is shown
that non of the reduced density matrix of any two subsystems depends on the
decoherence parameter irrespective of the channel used. That is, the reduced
density matrices of the subsystems behave as closed systems for which the
two-tangles are always zero. This means that no entanglement exists between
two subsystems. In other words, the entanglement resource cannot be utilized
by any two observers without the cooperation of the third one. Under the
influence of phase damping and phase flip channels, the one-tangles of the
accelerated observers are indistinguishable for the whole range of
acceleration and decoherence parameters. Similarly, in the limit of infinite
acceleration all the three one-tangles are equal which shows that all the
subsystems are equally entangled. On the other hand, the situation is
different when the system is coupled to a bit flip channel. That is, the
equality between any two one-tangles depends on which qubit is coupled to the
channel. In other words, the share of entanglement between subsystems is
dependent on the coupling of a particular qubit to the channel. This asymmetry
of the one tangles can be used to identify the frame of the observer coupled
to the noisy environment. Also, under no circumstances the three one-tangles
become equal for the whole range of decoherence parameter, except for a couple
of values, when the system is coupled to a bit flip channel. In the case of
phase damping channel, no sudden death of any one-tangle happens under local
coupling of any qubit with the channel. However, it goes to zero when the
channel is fully decohered. The sudden death of one-tangles may happen when
the system interacts with collective phase damping channel. Under the action
of phase flip channel the one-tangle sudden death occurs for every
acceleration and then a sudden rebirth of the one-tangle that increases as the
decoherence parameter increases. In the case of bit flip channel, the
one-tangle always survives and the effect of decoherence may be ignored in the
range of large acceleration. The sudden death of $\pi$-tangles may or may not
happens under the action of phase damping channel. Whereas it's sudden death
cannot be avoided when the system is influenced by phase flip channel. The
$\pi$-tangle is never lost when the system is coupled to a bit flip channel.
Which means that the entanglement of tripartite \textit{GHZ} state is robust
against bit flip noise. The fact that both one-tangles and $\pi$-tangle are
never lost in bit flip channel may be useful for faithful communication in
noninertial frames. Finally, it needs to be pointed out that the presence of a
noisy environment does not violate the CKW inequality \cite{Coffman}
$\mathcal{N}_{AB}^{2}+\mathcal{N}_{AC}^{2}\leq\mathcal{N}_{A(BC)}^{2}$ for
\textit{GHZ} initial state in the noninertial frames.

\end{document}